\documentclass[11pt]{article}
\usepackage{graphicx}
\usepackage{amsmath}
\usepackage{amssymb}
\usepackage{amsfonts}
\usepackage{caption}
\usepackage{mathtools}
\usepackage{color}
\usepackage[utf8]{inputenc}
\usepackage[english]{babel}
\usepackage[table]{xcolor}

\usepackage{array}
\usepackage{cellspace}
\usepackage[left=3cm,right=3cm, top=3cm, bottom=3cm]{geometry}
\newcolumntype{L}[1]{>{\raggedright\arraybackslash} m{#1} }
\usepackage[breaklinks=true,colorlinks=true,linkcolor=blue,urlcolor=blue,citecolor=blue]{hyperref}

\begin{document}

\title{\textbf {Deflection of light due to spheroidal oblate static objects}}
  \author{\textsc{Ranchhaigiri Brahma }\\ \textsl{Department of Physics, Assam University, Silchar-788011, India}\\ Email: \href{mailto: rbrahma084@gmail.com}{\textit{ rbrahma084@gmail.com}}\\
  \textsc{A.K. Sen  }\\ \textsl{Department of Physics, Assam University, Silchar-788011, India}\\ Email: \href{mailto: asokesen@yahoo.com}{\textit{ asokesen@yahoo.com}}}
\date{  }
\maketitle

\begin{abstract}
Deflection of light due to massive objects  was predicted by Einstein in his General Theory of Relativity. This deflection of light has been calculated by many researchers in past, for spherically symmetric objects. But, in reality, most of these gravitating objects are not spherical instead they are ellipsoidal  ( oblate) in shape. The objective of the present  work is to study theoretically the effect of this ellipticity on the trajectory of a light ray. Here, we obtain a converging series expression for the deflection of a light ray due to an ellipsoidal gravitating object, characterised by an ellipticity parameter.  As a  boundary condition, by setting the ellipticity  parameter to be equal to zero,  we get back the same expression for  deflection  as due to  Schwarzschild object.   It is also found that the additional contribution in deflection angle due to this ellipticity  though small, but  could be typically higher than the  similar contribution  caused by  the rotation of a celestial object.  Therefore for a precise estimate of the deflection due to a celestial object, the calculations presented here would be useful. 
\end{abstract}

\vspace{2cm}
\noindent{\it Keywords}: Schwarzschild-like solution, Oblate mass, Gravitational deflection, Null geodesics, Hamilton-Jacobi Equation

\section{Introduction}
According to Einstein's General Theory of Relativity, Gravitation is the manifestation of space-time geometry due to mass and energy \cite{1}. As a consequence of such geometry of space-time, the deflection of light due to a massive object was obtained by Einstein himself in 1915 \cite{2}. Einstein's prediction was successfully tested for the first time in 1919 by Eddington \cite{3}. In 1916 K. Schwarzschild obtained the exact solution of the Einstein's {\it Field Equation} for the curvature of space-time due to spherically symmetric distribution of mass \cite{4}. Reissner \& Nordström (1916-1918) \cite{5} obtained such a solution for a charged spherical massive object. After 47 years R.P. Kerr in 1963 obtained the exact solution for the rotating spherical mass which is given in terms of  Kerr line element \cite{6} and later, Newman (1965) \cite{7} obtained the  corresponding solution for  rotating-charged spherical mass which is known as Kerr-Newman line element. After Einstein calculated the expression for deflection of light due to the gravitational field  of static spherical mass, other researchers subsequently extended these calculations   to find the expressions for deflections under different space-time geometries.  Below, we go through some relevant work.

Hagihara (1930) \cite{8} studied the trajectory of light in the Schwarzschild  field and obtained the Hamilton-Jacobi equation for the said field. In 1936 Einstein investigated the lens like nature of the stars that bend the trajectory of the light \cite{9}. Later, Darwin (1959) \cite{10} calculated the deflection of light in strong gravitational field regime and obtained the result in logarithmic form. Later, Refsdal (1964) \cite{11} examined in detail the gravitational deflection of light and based on these suggested a method to obtain the mass of the gravitating object. Also Liebes, Jr. (1964) \cite{12} analysed the gravitational lensing and studied the properties of image formation due to such lensing. 
Luminet (1979) \cite{13} worked on the effect of strong gravitational  field on the trajectories of  light rays and the appearance of images due to such strong field.
Image formation due to large deflection angle has been studied by Ohanian (1987) \cite{14} and it was shown that under strong field,  the deflection angles of light can be expressed in terms of  elliptical integrals. 
Further, the effect of Schwarzschild black hole in strong field regime has been also analysed by Virbhadra \& Ellis (2000) \cite{15} using the null geodesic approach. The same authors  also studied the image formations around the optic axis of lens geometry. Then, for different black hole geometries, Bozza (2002) \cite{16} studied the effect of strong gravitational field and extended the expression under strong deflection limit, so as to allow the  complete capture of the photons by the black hole. Systematic study on the deflection angle of the light due to static and spherically symmetric objects have been reported by Keeton \& Petters (2005) \cite{17}. The authors showed that under weak deflection limit, the expression for deflection of light can be expanded in the Taylor series in order to have higher-order terms beyond the standard expression. Also, Iyer \& Petters (2007) \cite{18} have worked on both the strong and weak field conditions for gravitational deflection of light. They have reformulated the smooth transition from the strong to weak deflection limits  and then compared their values with numerically integrated ones. 
Bozza (2010) \cite{19} reviewed the theoretical development in gravitational deflection due to static and spherically symmetric fields both under strong and weak conditions. The author further extended the studies to include the effect of rotation of the gravitating object, on the trajectory of light ray. 
More recently, Huang (2017) \cite{20} has revisited the derivations for the deflection of light coming from celestial sources.

For a rotating spherical mass, Carter (1968) \cite{21} studied the geodesic equations in Kerr geometry . Later, Bray (1986)  \cite{22} obtained an expression for the trajectory of a light ray in the  Kerr field.  
Again, Sari (1996) \cite{23} worked on the deflection of light due to a slowly rotating spherical mass.
More recently, Bozza (2003)  \cite{24} calculated the lensing formula for Kerr black hole when the trajectory of the light  is very close to the equatorial plane. A more exact expression  for bending angle in the equatorial plane has obtained by Iyer \& Hansen (2009) \cite{25}  and showed that bending of light propagating in prograde direction is greater than that in the retrograde direction.
Aazami \textit{et. al.} (2011) \cite{26} further studied the trajectories of light rays in equatorial plane and in the field perpendicular to it  for Kerr geometry. In addition, Kraniotis (2011) \cite{27} obtained an equation of motion of light in Kerr gravitational field for the arbitrary inclination to the equatorial plane in presence of cosmological constant ($\Lambda$). Also, Chakraborty \& Sen (2017) \cite{28} worked on the deflection of light in Kerr geometry propagating slightly above the equatorial plane. 

Further, Hasse \& Pelrick (2006) \cite{29} studied the image formation in Kerr-Newman geometry and Kraniotis (2014) \cite{30} obtained an analytic solution for the deflection angle. Then, a series expression for the deflection angle due to the Kerr-Newman object has been obtained by Chakraborty \& Sen (2015)  \cite{31}.

Besides the standard null geodesic approach, some researchers have calculated the gravitational deflection of light by considering the curved space-time structure as equivalent to some optical medium. The concept of such an approach was given by Balazs (1958) \cite{32}. Using such a method, Sen (2010) \cite{33} obtained a more exact expression for deflection angle without any approximation for Schwarzschild geometry. Furthermore, Roy \& Sen (2015) \cite{34} calculated the deflection angle in the Kerr field using the same optical medium method.

All the above literature is based on the assumption that the gravitating objects are spherically symmetric, whereas most of the celestial objects are ellipsoidal (oblate) which may be because of their rotation or other factors. Therefore, understanding the significance of the non-spherical shape of the objects on the trajectories of light rays is important for precise calculation.
In 1959, Eroz \& Rosen \cite{35} obtained an exterior solution of the Einstein field equation for non-spherical deformed mass, for the first time which is known as Eroz-Rosen (E-R) metric. Then this solution was further investigated by some other authors in  \cite{36} \cite{37} \cite{38} \cite{39}. Hartle \& Thorne (1968) \cite{40} obtained the general solutions for slowly rotating axis-symmetric object. Besides these, a good number of cases  with space-time metrices for the mass with multiple moments has been discussed by some recent authors\cite{41} \cite{42} \cite{43}. Nikouravan (2011) \cite{44} obtained the Schwarzschild-like line element for space-time geometries due to static spheroidal oblate mass by performing oblate coordinate transformation in presence of mass.
Using this line element, Nikouravan \& Rawal (2013) \cite{45} studied the possible role of oblateness on the  the gravitational deflection  of light. On the other hand, Bini \& \textit{et.al.} (2013) \cite{46} investigated the trajectories of light rays in Eroz-Rosen space-time and obtained deflection angle of light in weak-field limit.

In the present work, we study the gravitational deflection of light using Nikouravan's solution in a more general way and obtain a series expression for the angle of deflection of light due to an oblate gravitating object. Here, we study the deflection angle for light  rays contained in the plane of ellipticity (plane perpendicular to the minor axis of an oblate) due to static gravitating objects. Further, we also compare the  additional contribution in deflection angle due to this ellipticity  with that which may be caused due to the typical rotation of a celestial body.

\section{Static Gravitational Field due to a Spheroidal oblate mass}
As stated above our work is limited to the static (\textit{i.e.} non-rotating) ellipsoidal (oblate) gravitating object, where we assumed such an object in space. Let the center of mass of this object be the origin of the Cartesian coordinate system in space. Let the ellipsoid has axes with lengths $\alpha$, $\beta$ and $\gamma$. If $\alpha =\beta >\gamma$, the object is known as a spheroidal oblate. The gravitational field due to such an oblate mass can be studied in an oblate coordinate system \cite{44}. Let us take the $z$-axis of our co-ordinate frame be oriented along $\gamma$ and   $x$ and $y$-axes be along $\alpha$ and $\beta$ respectively.  Thus one can write :  $x =\sqrt{r^2+a^2} \sin \theta \cos \phi,\,\, y =\sqrt{r^2+a^2} \sin \theta \sin \phi,\,\, z =r\cos \theta$, where $r$ is the radial coordinate, $\theta$ and $\phi$ are the angular coordinates measured from $z$ -axis and $y$ -axis  respectively. The parameter $a=\sqrt{\alpha^2-\gamma^2}$ is known as the linear eccentricity of the ellipsoid. \\\\
The line element obtained by Nikouravan (2011) \cite{44} for the space-time geometry due to the spheroidal oblate object is:

\begin{align}
ds^2=\left(1-\frac{2m}{r}\right)c^2dt^2-\frac{1}{\left(1-\frac{2m}{r}\right)}\left[\frac{r^2+a^2\cos^2 \theta}{r^2+a^2}\right]dr^2-&\left(r^2+a^2\cos^2\theta\right)d\theta^2\nonumber\\
&-\left(r^2+a^2\right)\sin^2\theta d\phi^2
\end{align}
where $2m=2MG/c^2 =r_g$ is the Schwarzschild radius; $m$ is known as Gravitational radius and  $M$ is the  actual mass of the object in physical unit.\\\\
From this line element (1), we have the metric elements as:
\begin{align}
g_{tt}&=\left(1-\frac{2m}{r}\right)\\
g_{rr}&=-\frac{1}{\left(1-\frac{2m}{r}\right)}\left[\frac{r^2+a^2\cos^2\theta}{r^2+a^2}\right]\\
g_{\theta \theta}&=-\left(r^2+a^2\cos^2\theta\right)\\
g_{\phi \phi}&=-\left(r^2+a^2\right)\sin^2\theta
\end{align}
In matrix form this can be written as:
\begin{align}
g_{ij}=\begin{pmatrix}
\left(1-\frac{2m}{r}\right) & 0 & 0 & 0\\[0.3em]
0 & -\frac{1}{\left(1-\frac{2m}{r}\right)}\left[\frac{r^2+a^2\cos^2\theta}{r^2+a^2}\right] & 0 & 0\\[0.3em]
0 & 0 & -\left(r^2+a^2\cos^2\theta\right) & 0\\[0.3em]
0 & 0 & 0 & -\left(r^2+a^2\right)\sin^2\theta
\end{pmatrix}
\end{align}
And its contravariant form can be obtained by using the relation, $g^{ij}=\frac{\text{co-factor\,} g_{ij}}{|g_{ij}|}$ as in \cite{34}.\\
Here, determinant of metric tensor $g_{ij}$ is obtained as below:
\begin{align}
|g_{ij}|&=\begin{vmatrix}
\left(1-\frac{2m}{r}\right) & 0 & 0 & 0\\[0.3em]
0 & -\frac{1}{\left(1-\frac{2m}{r}\right)}\left[\frac{r^2+a^2\cos^2\theta}{r^2+a^2}\right] & 0 & 0\\[0.3em]
0 & 0 & -\left(r^2+a^2\cos^2\theta\right) & 0\\[0.3em]
0 & 0 & 0 & -\left(r^2+a^2\right)\sin^2\theta
\end{vmatrix}\\ \nonumber\\ \nonumber
&=\left(1-\frac{2m}{r}\right)\begin{vmatrix}
-\frac{1}{\left(1-\frac{2m}{r}\right)}\left[\frac{r^2+a^2\cos^2\theta}{r^2+a^2}\right] & 0 & 0\\[0.3em]
0 & -\left(r^2+a^2\cos^2\theta\right) & 0\\[0.3em]
0 & 0 & -\left(r^2+a^2\right)\sin^2\theta
\end{vmatrix}\\ \nonumber\\ \nonumber
&=-\left(r^2+a^2\cos^2\theta\right)^2\sin^2\theta
\end{align}
The co-factor of the metric $g_{ij}$ can be obtained by using the formula that for a matrix $A$ having elements $A_{ij}$, the co-factor of the elements $A_{ij}$ is  $=\left(-1\right)^{i+j}\big|M_{ij}\big|$, where $i, j$ denote the row number and column number of a particular element $A_{ij}$ of the matrix $A$ and $|M_{ij}|$ is the determinant of the minor for $(ij)^{th}$ element. The non-zero co-factors of the elements of $g_{ij}$ are as below:
\begin{align*}
\text{co-factor of}\,\, g_{tt} &=-\frac{\left(r^2+a^2\cos^2\theta\right)^2\sin^2\theta}{\left(1-\frac{2m}{r}\right)}\\
\text{co-factor of}\,\, g_{rr} &= \left(1-\frac{2m}{r}\right)\left(r^2+a^2\cos^2\theta\right)\left(r^2+a^2\right)\sin^2\theta\\
\text{co-factor of}\,\, g_{\theta\theta} &= \left(r^2+a^2\cos^2\theta\right)\sin^2\theta\\
\text{co-factor of}\,\, g_{\phi\phi}&=\frac{\left(r^2+a^2\cos^2\theta\right)^2}{r^2+a^2}
\end{align*}
Therefore, we have the contravariant form of the metric as:
(using $g^{ij}=\frac{\text{co-factor\,} g_{ij}}{|g_{ij}|}$),
\begin{align}
g^{ij}=\begin{pmatrix}
\frac{1}{\left(1-\frac{2m}{r}\right)} & 0 & 0 & 0\\[0.3em]
0 & -\frac{\left(1-\frac{2m}{r}\right)\left(r^2+a^2\right)}{r^2+a^2\cos^2\theta} & 0 & 0\\[0.3em]
0 & 0 & -\frac{1}{r^2+a^2\cos^2\theta} & 0\\[0.3em]
0 & 0 & 0 & -\frac{1}{\left(r^2+a^2\right)\sin^2\theta}
\end{pmatrix}
\end{align}
\subsection{General Hamilton-Jacobi Equation}
For the study of trajectory of light, we apply the null geodesic approach. Before going to the main formulation, let us work-out some important tasks. First of all, in the general theory of relativity, Lagrangian (per unit mass) of a system is given by \cite{5}\cite{31}
\begin{align}
f=\frac{1}{2}g_{ij}\dot{x}^{i}\dot{x}^{j}=\frac{1}{2}\left[g_{tt}c^{2} \dot{t}^2-g_{rr} \dot{r}^2-g_{\theta\theta}\dot{\theta}^2-g_{\phi \phi}\dot{\phi}^2\right]
\end{align}
where, dot over $t$, $r$, $\theta$ and $\phi$ denotes the differentiation with respect to an affine parameter $\lambda$ along the geodesic. (Affine parameter is defined along the path of the geodesic \cite{1}. For time like geodesics, it is related to proper time and for space like geodesics, it is related to proper distance.)\\\\
Now, from the Euler-Lagrange equation, we have, 
\begin{align*}
\frac{d}{d\lambda}\left(\frac{\partial f}{\partial\dot{x}^i}\right)-\frac{\partial f}{\partial x^i}=0
\end{align*}
As the Lagrangian is independent of $t$ and $\phi$ coordinates, we obtain  below the following equations:
\begin{align}
\frac{d}{d\lambda}\left(\frac{\partial f}{c \partial \dot{t}}\right)&-\frac{\partial f}{c \partial t}=0\Rightarrow\frac{d}{d\lambda}\left(\frac{\partial f}{c \partial\dot{t}}\right)-0=0\nonumber\\
\Rightarrow\left(\frac{\partial f}{c \partial\dot{t}}\right)&=\left(1-\frac{2m}{r}\right) c\dot{t}=E\text{  (a constant)}\\
\text{and     }\frac{d}{d\lambda}\left(\frac{\partial f}{\partial\dot{\phi}}\right)&-\frac{\partial f}{\partial \phi}=0\Rightarrow\frac{d}{d\lambda}\left(\frac{\partial f}{\partial\dot{\phi}}\right)-0=0\nonumber\\
\Rightarrow\left(\frac{\partial f}{\partial\dot{\phi}}\right)&=\left(r^2+a^2\right)\sin^2 \theta\dot{\phi}=L\text{  (a constant)}
\end{align}
Since, $E$ and $L$ are the constants of motion, hence they are the conserved quantities in the respective coordinate system. These two constants $E$ and $L$ respectively represent the energy and the angular momentum (per unit mass) around the short $\gamma$ axis.\\
And the corresponding canonical momenta for other two coordinates can be obtained from the relation $p_{i}=\frac{\partial f}{\partial\dot{x}^{i}}=g_{ij}\dot{x}^{j}$ as below: 
\begin{align}
p_{r}&=\frac{1}{\left(1-\frac{2m}{r}\right)}\left[\frac{r^2+a^2\cos^2\theta}{r^2+a^2}\right]\dot{r}\\
\text{and      }p_{\theta}&=\left(r^2+a^2\cos^2\theta\right)\dot{\theta}
\end{align}
Again, the Hamiltonian of the system is given by \cite{5}\cite{21}
\begin{align}
H(x^{i}, p_{j})&=p_{i}\dot{x}^{i}(p_{j})-f(x^{i}, \dot{x}^{i})=\frac{1}{2}g^{ij}p_{i}p_{j}
\end{align}
and the Hamilton-Jacobi equation with solution $S\equiv S(x^{i},\lambda)$  is :
\begin{align}
H\left(x^{i},\frac{\partial S}{\partial x^{i}}\right)+\frac{\partial S}{\partial \lambda}=0
\end{align}
As the Hamiltonian is a function of coordinate and momentum, and  $\dot{x}^{i}=g^{ij}p_{j}$,  so here we have:
\begin{align}
\frac{\partial S}{\partial x^{i}}&=p_{i}\\
\text{and}\hspace*{1cm}
\dot{x^{i}}&=\frac{dx^{i}}{d\lambda}=g^{ij}\frac{\partial S}{\partial x^{i}}
\end{align}
Thus in the above, we have seen that the Lagrangian of the ellipsoidal field is independent of $t$ and $\phi$, and so the momentum associated with these coordinates are constants of motion. The null geodesic action with such constants of motion can be written as ( considering  $r$ and $\theta$ as separable coordinates) :
\begin{align}
S=-Et+L\phi+S_{r}(r)+S_{\theta}(\theta)
\end{align}
where $S_r$ and $S_{\theta}$ are the action associated with and only dependent on $r$ and $\theta$ coordinates respectively. The Hamilton-Jacobi equation with such action for ellipsoidal field can be obtained as follows:
\begin{small}
\begin{align}
&\frac{1}{2}g^{ij}\frac{\partial S}{\partial x^{i}}\frac{\partial S}{\partial x^{j}}+\frac{\partial S}{\partial\lambda}=0\nonumber\\
\text{or}\,\,\,\,&\frac{1}{2}\left[g^{tt}\left(\frac{\partial S}{c\partial t}\right)^2+g^{rr}\left(\frac{\partial S}{\partial r}\right)^2+g^{\theta\theta}\left(\frac{\partial S}{\partial \theta}\right)^2+g^{\phi\phi}\left(\frac{\partial S}{\partial \phi}\right)^2\right]=0\nonumber\\
\text{or}\,\,\,\,&-\frac{E^2}{c^2\left(1-\frac{2m}{r}\right)}+\left[\frac{\left(1-\frac{2m}{r}\right)(r^2+a^2)}{r^2+a^2\cos^2\theta}\right]\left(\frac{\partial S_r}{\partial r}\right)^2+\frac{1}{r^2+a^2\cos^2\theta}\left(\frac{\partial S_{\theta}}{\partial \theta}\right)^2+\frac{L^2}{\left(r^2+a^2\right)\sin^2\theta}=0\nonumber\\
\text{or}\,\,\,\,&-\frac{E^2(r^2+a^2\cos^2\theta)}{c^2\left(1-\frac{2m}{r}\right)}+\left[\left(1-\frac{2m}{r}\right)(r^2+a^2)\right]\left(\frac{\partial S_r}{\partial r}\right)^2+\left(\frac{\partial S_{\theta}}{\partial \theta}\right)^2+\frac{(r^2+a^2\cos^2\theta)L^2}{\left(r^2+a^2\right)\sin^2\theta}=0\nonumber\\
\text{or}\,\,\,\,&\left[\left(1-\frac{2m}{r}\right)(r^2+a^2)\right]\left(\frac{\partial S_r}{\partial r}\right)^2+\left(\frac{\partial S_{\theta}}{\partial \theta}\right)^2-F\left(r,\theta\right)=0
\end{align}
\end{small}
where,
\begin{align}
F\left(r,\theta\right)&=\left(r^2+a^2\cos^2\theta\right)\left[\frac{E^2}{c^2\left(1-\frac{2m}{r}\right)}-\frac{L^2}{\left(r^2+a^2\right)\sin^2\theta}\right]
\end{align}
Eqn.(19) is the general Hamilton-Jacobi equation corresponding to a light  ray  moving in the  gravitational field of an oblate object.

\section{Trajectory of light in the plane of oblateness}
Let the entire trajectory of the light is contained in the plane of oblateness i.e. the plane with $\theta=\pi/2$ (which is also the x-y plane) . Thus, we have,
\begin{align}
F\left(r,\pi/2\right)&=r^2\left[\frac{E^2}{c^2\left(1-\frac{2m}{r}\right)}-\frac{L^2}{\left(r^2+a^2\right)}\right]\\
L&=\left(r^2+a^2\right)\dot{\phi}\\
\text{and}\,\,\,\,\left(\frac{\partial S_{\theta}}{\partial \theta}\right)^2&=0
\end{align}
Therefore, the Hamilton-Jacobi equation (19) will be as follows:
\begin{align}
\left[\left(1-\frac{2m}{r}\right)(r^2+a^2)\right]\left(\frac{\partial S_r}{\partial r}\right)^2&=r^2\left[\frac{E^2}{c^2\left(1-\frac{2m}{r}\right)}-\frac{L^2}{\left(r^2+a^2\right)}\right]\nonumber\\
\text{or          }\left(\frac{\partial S_r}{\partial r}\right)^2&=r^2\left[\left(1-\frac{2m}{r}\right)(r^2+a^2)\right]^{-1}\left[\frac{E^2}{c^2\left(1-\frac{2m}{r}\right)}-\frac{L^2}{\left(r^2+a^2\right)}\right]
\end{align}
And hence we obtained radial geodesic from Eqn. (17) as:
\begin{align}
\dot{r}^2&=\left(\frac{dr}{d\lambda}\right)^2=\left(g^{rr}\frac{\partial S_r}{\partial r}\right)^2\nonumber\\
&=\left(1-\frac{2m}{r}\right)^2\left[\frac{r^2+a^2}{r^2}\right]^2r^2\left[\left(1-\frac{2m}{r}\right)(r^2+a^2)\right]^{-1}\left[\frac{E^2}{c^2\left(1-\frac{2m}{r}\right)}-\frac{L^2}{\left(r^2+a^2\right)}\right]\nonumber\\
&=\left(1-\frac{2m}{r}\right)\left[\frac{r^2+a^2}{r^2}\right]\left[\frac{E^2}{c^2\left(1-\frac{2m}{r}\right)}-\frac{L^2}{\left(r^2+a^2\right)}\right]\nonumber\\
&=L^2\left(1-\frac{2m}{r}\right)\left[\frac{r^2+a^2}{r^2}\right]\left[\frac{E^2/c^2L^2}{\left(1-\frac{2m}{r}\right)}-\frac{1}{\left(r^2+a^2\right)}\right]\\
\text{or \,\,\,\,\,}\dot{r}&=\pm L\sqrt{\frac{\frac{1}{b^2}-\frac{\left(1-\frac{2m}{r}\right)}{\left(r^2+a^2\right)}}{\left(1-\frac{2m}{r}\right)\left(1-\frac{2m}{r}\right)^{-1}\left[\frac{r^2+a^2}{r^2}\right]^{-1}}}\nonumber\\
&=\pm L\sqrt{\frac{\frac{1}{b^2}-\frac{\left(1-\frac{2m}{r}\right)}{\left(r^2+a^2\right)}}{\left(\frac{r^2}{r^2+a^2}\right)}}
\end{align}
where the term denoted by $b=cL/E$ in the above equation is known as the apparent impact parameter or simply impact parameter \cite{47}, in which $L$ and $E$ are angular momentum and energy of the photon respectively, as stated earlier. Since the light is propagating with momentum $p$ and energy $E$ having speed $c$, we can write $E=pc\,\,\, \text{and also  } L=pb\Rightarrow b=cL/E$. This impact parameter $(b)$ can be identified as the perpendicular distance between the centre of the gravitating object and the trajectory of the light ray. If the light is travelling parallel to the $x-$direction and contained in the $x-y$ plane, then this impact parameter is simply represented by the $y-$ coordinate of light ray had there been no gravitational field.                
\subsection{Closest Approach}
The closest approach of the light ray $(r_0)$ is the distance between the center of gravitating object and the trajectory of the light when it is closest to the gravitating object. It is different from the impact parameter $(b)$, as the impact parameter is the measure of the distance between center of the object and the trajectory of light when there is no gravitational field. The relation between $b$ and $r_0$ can be obtained by taking the condition for the extremum value of Eqn. (26). That is, if the light ray approaches to the closest distance $r=r_0$ ( at affine parameter $\lambda=\lambda_0$); then $\dot{r}=0$ and from the above Eqn. (26) we can write:
\begin{align}
\frac{1}{b^2}=\frac{1-\frac{2m}{r_0}}{r_0^2+a^2}
\end{align}
\begin{align}
\text{or\,\,\,\,}r_0^3-r_0(b^2-a^2)+2mb^2=0
\end{align}
The above Eqn. (28) is a cubic equation in $r_0$.
The solution of this cubic equation can be obtained by using trigonometric identity:
\begin{align}
4\cos^3 \vartheta-3\cos \vartheta-\cos (3\vartheta)=0
\end{align}
For that, we considered $r_0=n\cos\vartheta $; where $n$ is a constant. Substituting this into Eqn. (28) we obtain:
\begin{align}
r_0^3-(b^2-a^2)r_0+2mb^2=n^3\cos^3 \vartheta-(b^2-a^2)n\cos\vartheta +2mb^2=0
\end{align}
Now, comparing Eqn. (29) and (30) we get:
\begin{align*}
\frac{4}{n^3}=\frac{3}{(b^2-a^2)n}=-\frac{\cos (3\vartheta)}{2mb^2}
\end{align*}
After simplification we can write:
\begin{align*}
n&=2\sqrt{\frac{(b^2-a^2)}{3}}\\
\text{and    } \cos (3\vartheta)&=-\frac{3(2mb^2)}{(b^2-a^2)n}\\
\Rightarrow \vartheta&=\frac{1}{3}\cos^{-1}\left(\frac{3(2mb^2)}{-(b^2-a^2)n}\right)=\frac{1}{3}\cos^{-1}\left(\frac{3(2mb^2)}{-2(b^2-a^2)}\sqrt{\frac{3}{(b^2-a^2)}}\right)
\end{align*}
Therefore, the root of the Eqn. (28) is,
\begin{align}
r_0&=n\cos \vartheta\nonumber\\
&=2\sqrt{\frac{(b^2-a^2)}{3}}\cos \left[\frac{1}{3}\cos^{-1}\left(\frac{3(2mb^2)}{-2(b^2-a^2)}\sqrt{\frac{3}{(b^2-a^2)}}\right)\right]
\end{align}
Since the value of $\cos (3\vartheta)$ remains unchanged by substituting $\vartheta_1=\vartheta+\frac{2\pi}{3}$ and $\vartheta_2=\vartheta+\frac{4\pi}{3}$ in the place of $\vartheta$, we get that the cubic equation (28) has three roots given by $n \cos \vartheta$, $n\cos \vartheta_1$ and $n\cos \vartheta_2$. These roots can be expressed in a single general form by considering $k$ as the number having values $0, 1, 2$ as below:
\begin{align}
r_0&=2\sqrt{\frac{b^2-a^2}{3}}\cos \left[\frac{1}{3}\cos^{-1}\left(\frac{3\times 2mb^2}{-2(b^2-a^2)}\sqrt{\frac{3}{b^2-a^2}}\right)+\frac{2\pi k}{3}\right]
\end{align}
We note that,  the light is not  captured while approaching the gravitating object and instead it has a point of bending at the largest value of $r_0$. This indicates the closest approach corresponds to  $r_0$ as the largest root\cite{47} of Eqn. (28). The largest value of $r_0$ can be obtained by putting $k=0$ in Eqn. (32). By plotting the above Eqn. (32) for $r_0$ with respect to $k$ from $0\longrightarrow 2$ we can find its largest value at $k=0$,  as shown in \textbf{Figure 1}.
\begin{figure}[h!]
\centering
\minipage{0.7\textwidth}
\includegraphics[width=\linewidth]{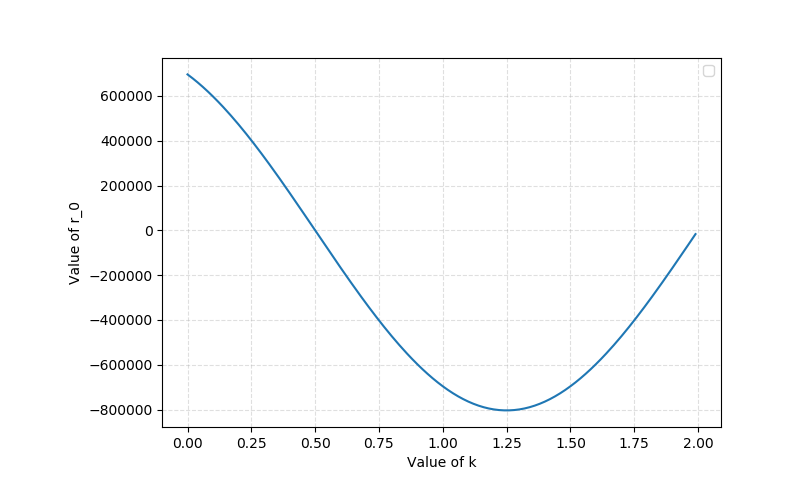}
\caption{\textit{Characteristic curve of $r_0$ showing largest value at $k=0$}}
\endminipage
\end{figure}

Hence, we obtain the expression for the closest approach of light as below:
\begin{align}
r_0&=2\sqrt{\frac{b^2-a^2}{3}}\cos \left[\frac{1}{3}\cos^{-1}\left(\frac{3\times 2mb^2}{-2(b^2-a^2)}\sqrt{\frac{3}{b^2-a^2}}\right)\right]\nonumber\\
&=2\sqrt{\frac{b^2-a^2}{3}}\cos\left[\frac{1}{3}\cos^{-1}\left(-\frac{3^{3/2}m}{b\left(1-\frac{a^2}{b^2}\right)^{3/2}}\right)\right]\nonumber\\
\text{or\,\,\,\,}\frac{r_0}{b_0}&=\frac{2}{\sqrt{3}}\cos\left[\frac{1}{3}\cos^{-1}\left(-\frac{3^{3/2}m}{b^{\star}}\right)\right]
\end{align}
Here, we put $\sqrt{b^2-a^2}=b\sqrt{1-\frac{a^2}{b^2}}=b_0$ and $b\left(1-\frac{a^2}{b^2}\right)^{3/2}=b^{\star}$.
This Eqn. (33) is the general expression for the closest approach in terms of impact parameter $(b)$.\\\\
For weak field condition,  we should have $\frac{m}{b^{\star}}\ll 1$.  Hence we can expand the right hand side of Eqn. (33) in Taylor series \cite{17} in the power of  $\frac{m}{b\left(1-\frac{a^2}{b^2}\right)^{3/2}}=\frac{m}{b^{\star}}$  about the point at $0$ as below:

\begin{align}
\frac{r_0}{b_0}&=1-\frac{m}{b^{\star}}-\frac{3}{2}\left(\frac{m}{b^{\star}}\right)-4\left(\frac{m}{b^{\star}}\right)+....\nonumber\\
\text{or\,\,\,\,}r_0&=b_0\left[1-\frac{m}{b^{\star}}-\frac{3}{2}\left(\frac{m}{b^{\star}}\right)-4\left(\frac{m}{b^{\star}}\right)+....\right]\nonumber\\
\text{or\,\,\,\,}r_0&=b_0-\frac{m}{\left(1-\frac{a^2}{b^2}\right)}-\frac{3}{2}\frac{m^2}{b\left(1-\frac{a^2}{b^2}\right)^{5/2}}-4\frac{m^3}{b^2\left(1-\frac{a^2}{b^2}\right)^{4}}+....\nonumber\\
\text{or\,\,\,\,}r_0&=b_0-\frac{m}{K}-\frac{3}{2}\frac{m^2}{bK^{5/2}}-4\frac{m^3}{b^2K^{4}}+....
\end{align}
where, we denoted $(1-\frac{a^2}{b^2})$ by $K$.
\subsection{Deflection Angle of Light}
As the light is travelling near the gravitating object, the trajectory of the light and the center of mass of the object makes a plane. Since the trajectory of light is in that plane, where $\theta=\pi/2$ as stated earlier, this can be described by the coordinate $(r,\phi)$ and the expression for the trajectory is, \cite{5}\cite{17}\cite{48}
\begin{align}
\Delta\phi =2\int_{r_0}^{\infty}\bigg|\frac{d\phi}{dr}\bigg|dr
\end{align}
The trajectory of light is a straight line in the absence of a gravitational field. Hence, the angle of deflection of the light due to gravitational effect is given by  \cite{5}\cite{17}\cite{33}\cite{48}

\begin{align}
\Delta \phi =2\int_{r_0}^{\infty}\bigg|\frac{d\phi}{dr}\bigg|dr-\pi
\end{align}
Now, from Eqn. (11) and (26) we have:
\begin{align}
\frac{d\phi}{dr}=\frac{1}{r^2+a^2}\left[\frac{\frac{r^2}{r^2+a^2}}{\frac{1}{b^2}-\frac{\left(1-\frac{2m}{r}\right)}{\left(r^2+a^2\right)}}\right]^{1/2}
\end{align}
Therefore, substituting Eqn. (37) into Eqn. (36) we obtain:   
\begin{align}
\Delta\phi&=2\int_{r_0}^{\infty}\frac{1}{r^2+a^2}\left[\frac{\frac{r^2}{r^2+a^2}}{\frac{1}{b^2}-\frac{\left(1-\frac{2m}{r}\right)}{\left(r^2+a^2\right)}}\right]^{1/2}dr-\pi\nonumber\\
&=2\int_{r_0}^{\infty}\frac{1}{r^2+a^2}\left[\frac{\frac{r^2}{r^2+a^2}}{\frac{1}{b^2}-\frac{1}{r^2+a^2}+\frac{2m}{r(r^2+a^2)}}\right]^{1/2}dr-\pi
\end{align}
For simplicity, let us assume a variable, $\xi^2=\frac{r_0^2+a^2}{r^2+a^2}$ and a constant, $h=\frac{m}{r_0}$. Then we write the following: 
\begin{align}
\xi^2&=\frac{r_0^2+a^2}{r^2+a^2}\hspace{1cm} \Rightarrow r^2+a^2=\frac{r_0^2+a^2}{\xi^2}\hspace{1cm}\Rightarrow r=\sqrt{\frac{r_0^2+a^2}{\xi^2}-a^2}
\end{align}
From Eqn. (39), we have:
\begin{align}
\frac{d}{d\xi}&(r^2+a^2)=\frac{d}{d\xi}\left(\frac{r_0^2+a^2}{\xi^2}\right)\hspace{1cm}\Rightarrow 2r\frac{dr}{d\xi}=-2(r_0^2+a^2)\frac{1}{\xi^3}\nonumber\\
\text{or\,\,\,\,}\frac{dr}{d\xi}&=-(r_0^2+a^2)\frac{1}{r\xi^3}=-(r_0^2+a^2)\frac{1}{\xi^3\sqrt{\frac{r_0^2+a^2}{\xi^2}-a^2}}\nonumber\\
&=-(r_0^2+a^2)\frac{1}{\xi^2\sqrt{r_0^2+a^2-\xi^2a^2}}\nonumber\\
\text{or\,\,\,\,}dr&=-\frac{(r_0^2+a^2)d\xi}{\xi^2\sqrt{r_0^2+a^2-\xi^2a^2}}
\end{align}
And now the limits will be as $r\longrightarrow \infty$ then $\xi\longrightarrow 0$ and as $r\longrightarrow r_0$ then $\xi\longrightarrow 1$. 
Substituting from Eqn. (39), (40) into Eqn.(38) we have,
\begin{align}
\Delta\phi &=-2\int_{1}^0\frac{\xi^2}{(r_0^2+a^2)}\left[\frac{\frac{r^2\xi^2}{r_0^2+a^2}}{\frac{1}{b^2}-\frac{\xi^2}{r_0^2+a^2}+\frac{2m\xi^2}{r(r_0^2+a^2)}}\right]^{1/2}\frac{(r_0^2+a^2)d\xi}{\xi^2\sqrt{r_0^2+a^2-\xi^2a^2}}-\pi\nonumber\\
&=2\int_{0}^1\left[\frac{\frac{r^2\xi^2}{r_0^2+a^2}}{\frac{1}{b^2}-\frac{\xi^2}{r_0^2+a^2}+\frac{2m\xi^2}{r(r_0^2+a^2)}}\right]^{1/2}\frac{d\xi}{\sqrt{r_0^2+a^2-\xi^2a^2}}-\pi\nonumber\\
&=2\int_{0}^1\left[\frac{\left(\frac{r_0^2+a^2}{\xi^2}-a^2\right)\frac{\xi^2}{r_0^2+a^2}}{\frac{1}{b^2}-\frac{\xi^2}{r_0^2+a^2}+\frac{2m\xi^2}{\sqrt{\frac{r_0^2+a^2}{\xi^2}-a^2}\,(r_0^2+a^2)}}\right]^{1/2}\frac{d\xi}{\sqrt{r_0^2+a^2-\xi^2a^2}}-\pi\nonumber\\
&=2\int_0^1\left[\frac{\frac{1}{\xi^2}(r_0^2+a^2-\xi^2a^2)\frac{\xi^2}{(r_0^2+a^2)}}{\frac{1-\frac{2m}{r_0}}{r_0^2+a^2}-\frac{\xi^2}{r_0^2+a^2}+\frac{2m\xi^3}{\sqrt{r_0^2+a^2-\xi^2a^2}(r_0^2+a^2)}}\right]^{1/2}\frac{d\xi}{\sqrt{r_0^2+a^2-\xi^2a^2}}-\pi\nonumber\\
&=2\int_0^1\frac{d\xi}{\sqrt{1-\frac{2m}{r_0}-\xi^2+\frac{2m\xi^3}{r_0\sqrt{1+\frac{a^2}{r_0^2}(1-\xi^2)}}}}-\pi\nonumber\\
&=2\int_0^1\frac{d\xi}{\sqrt{1-\xi^2-2h+2h\xi^3P(\xi)}}-\pi
\end{align}
where, $P(\xi)=\frac{1}{\sqrt{1+\frac{a^2}{r_0^2}(1-\xi^2)}}$. 
If $a=0$ then, $P(\xi)=1$ and if $a\neq 0$ then, $P(\xi)\neq 1 $. This clearly indicates that, Eqn. (41) reduces to the gravitational deflection of light due to spherically symmetric static objects at $a=0$. Therefore, the function $P(\xi)$ contains the information about the contribution of ellipsoidal shape ( or ellipticity)  on the gravitational deflection of light. The Eqn.(41) is the general expression for the deflection of light due to ellipsoidal gravitating object.
\subsection{Weak Field condition}
For weak field conditions, the trajectory of the light is assumed to be far from the photon sphere and hence closest approach $(r_0)$ is beyond the photon radius denoted by $(r_{ph})$. In such condition, $h=\frac{m}{r_0}< 1/3$ \cite{17}. Let, $Q(\xi)=\left[1-\xi^2-2h+2h\xi^3P(\xi)\right]^{-1/2}$and then expanding $Q(\xi)$ in Taylor series form in the power of $h$ we have:
\begin{align}
Q(\xi)&=Q(0)+Q'(0)h+Q''(0)\frac{h^2}{2!}+...
\end{align}
where, $(')$ denotes the derivative with respect to $(h)$. 
So, we obtain a series expression for deflection angle from Eqn. (41) as below:
\begin{align}
\Delta \phi&=2\int_0^1\left[Q(0)+Q'(0)h+Q''(0)\frac{h^2}{2!}+...\right]d\xi-\pi\nonumber\\
&=\Delta\phi_0+\Delta\phi_1+\Delta\phi_2+...-\pi
\end{align}
Evaluating Eqn. (43) term by term we get:
\begin{align}
\Delta \phi_0=2\int_0^1 Q(0)d\xi=2\int_0^1\frac{d\xi}{\sqrt{1-\xi^2}}=\pi
\end{align}
Similarly, for the first and second order terms we have : 
\begin{align}
\Delta \phi_1&=2h\int_0^1Q'(0)d\xi\nonumber\\
&=2h\int_0^1\frac{d}{dh}\left[1-\xi^2-2h+2h\xi^3P(\xi)\right]^{-1/2}\Big|_{h=0}d\xi\nonumber\\
&=2h\int_0^1\frac{1-\xi^3\left[1-\frac{a^2}{2r_0^2}\left(1-\xi^2\right)\right]}{\left(1-\xi^2\right)^{3/2}}d\xi\nonumber\\
&=2h\left(2+\frac{a^2}{3r_0^2}\right)
\end{align}
\begin{align}
\text{and}\hspace{1cm}\Delta \phi_2&=h^2\int_0^1Q''(0)d\xi\nonumber\\
&=h^2\int_0^1\frac{d''}{dh}\left[1-\xi^2-2h+2h\xi^3P(\xi)\right]^{-1/2}\Big|_{h=0}d\xi\nonumber\\
&=3h^2\int_0^1\frac{\left(1-\xi^3\left[1-\frac{a^2}{2r_0^2}\left(1-\xi^2\right)+\frac{3}{8}\left\{\frac{a^2}{r_0^2}(1-\xi^2)\right\}^2\right]\right)^2}{(1-\xi^2)^{5/2}}d\xi\nonumber\\
&=h^2\left[-4+\frac{15\pi}{4}+\frac{81a^8\pi}{32768r_0^8}-\frac{45a^6\pi}{2048r_0^6}+\frac{3a^4\left(-16+5\pi\right)}{32r_0^4}+\frac{3a^2\left(-32+15\pi\right)}{16r_0^2}\right]
\end{align}

Hence, from Eqn. (43), (44), (45) and (46), we have obtained an expression for the  deflection of light  as:
\begin{align}
\Delta \phi&=2h\left(2+\frac{a^2}{3r_0^2}\right)+h^2\Big[-4+\frac{15\pi}{4}+\frac{81a^8\pi}{32768r_0^8}-\frac{45a^6\pi}{2048r_0^6}\nonumber\\
&\hspace{2.5cm}+\frac{3a^4\left(-16+5\pi\right)}{32r_0^4}+\frac{3a^2\left(-32+15\pi\right)}{16r_0^2}\Big]+\mathcal{O}\left(h\right)^3
\end{align}
This Eqn. (47) is the expression for the gravitational deflection of light due to an ellipsoidal oblate object under weak field condition. Here, it is also observed that, when $a=0$ the Eqn. (47) reduces to the standard series expression for the deflection angle of light due to Schwarzschild mass as obtained by Keeton \& Petters (2005) \cite{17} and Sen (2010)\cite{33}.

\section{Discussion of Results}
To understand the important contribution of ellipticity of  gravitating object on the deflection of light ray, let us consider Sun as our test object.  The Sun has Schwarzschild radius approximately $2m=r_g=3\,\,km$, rotation parameter $\alpha_{rot}=1.69749 \,\,km$ and linear eccentricity $a=2951.6 \,\,km$. Assuming the closest approach of light ray equal to its equatorial radius $R=695700 \,\,km$, we calculated the gravitational deflection angle of light for rotating mass both in prograde and retrograde direction. For that calculation, we used the expression obtained by Chakraborty \& Sen (2015) \cite{31} considering first-order term and putting charge parameter as zero.  Then, we also calculated the angle of deflection due to ellipsoidal mass by using our Eqn. (47) and compared both the values with the standard Schwarzschild gravitational deflection. From that, it is observed that the angle of deflection due to rotating mass is slightly greater in the retrograde direction and smaller in the prograde direction compared to Schwarzschild deflection. Again, we also observe that, the deflection angle due to Sun as an oblate object is larger than the Schwarzschild deflection as well as Kerr deflection (Numerical values are shown in \textbf{Table 1}).  
\begin{table}[h!]
\small
\caption{Comparison of contributions of rotation and oblateness on deflection of light. }
\label{tab:table2}\centering
\renewcommand\arraystretch{1.3}
\begin{tabular}{ | L{\dimexpr 0.07\linewidth-2\tabcolsep}| L{\dimexpr 0.27\linewidth-2\tabcolsep} | L{\dimexpr 0.17\linewidth-2\tabcolsep}  | L{\dimexpr 0.17\linewidth-2\tabcolsep}| L{\dimexpr 0.24\linewidth-2\tabcolsep} |}  
\hline
\textit{Sl. No.} & \textit{Geometry} & \begin{center} $\Delta\phi$ 

\textit{(arc sec.)}

\end{center} & 
\begin{center}Schwarzschild
deflection

\textit{(arc sec.)}

$\Delta\phi_s$
\end{center} &  \begin{center}
$\Delta\phi-\Delta\phi_s$
\end{center}
\\ \hline
1 & Kerr (prograde) \cite{31} & $1.7789092$  &  & $-2.411550\times10^{-6}$\\ 
2 & Kerr (retrograde) \cite{31} & $1.7789140$ &  $1.7789116$ & $2.411603\times10^{-6}$\\ 
3 & Ellipsoidal (Eqn. (47)) & $1.7789169$ &  & $5.336716\times10^{-6}$\\ 
 & (present work)&  &  & \\ \hline
\end{tabular}
\end{table}
From the \textbf{Table 1}, we can see that the contribution of oblateness of the Sun in the deflection of light is sufficiently larger than that due to the rotation parameter. This result showing the contribution of the oblateness (or ellipticity in general) of a gravitating object in the trajectory of light ray tells  us clearly that, this is an important and  sizeable contribution. So it should be included for all precise calculation  of gravitational deflection in future. However, such contribution will be obviously different for objects having different values of oblateness parameter $(a)$  or ellipticity. 


\section{Conclusions}
In this work, we have derived an expression for deflection of light due to static oblate mass in a converging series form.  The expression reduces to Schwarzschild deflection if the ellipticity  parameter is set to zero. After calculating the numerical values of deflection for Sun considering separately its rotation and oblateness, we find the contribution due to oblateness is greater than that due to the rotation. Therefore, the contribution of oblateness can't be neglected  to have accurate values of deflection in all such future calculations. 

\subsection*{Acknowledgements}
We would like to thank Prof. A. Deshamukhya and Prof. B. Indrajit Sharma (HoD, Physics) Assam University, Silchar, India for their constant support and encouragement.

\end{document}